\documentclass[iop,usenatbib]{emulateapj}

\def\1{\c{c}}
\def\2{\c{C}}
\def\3{\.{I}}
\def\4{\"{a}}
\def\5{{\i}}
\def\6{$\beta$}
\def\7{\"{o}}
\def\8{\"{O}}
\def\9{\c{s}}
\def\0{\c{S}}
\def\*{\"{u}}
\def\?{\"{U}}
\def\;{\u{g}}
\def\:{\u{G}}
\usepackage{color}
\usepackage{graphicx}
\usepackage{epsfig}

\slugcomment{Published in the Astrophysical Journal (08 Jun 2017).}

\shorttitle{Recombining Plasma \& Gamma-ray Emission in 3C 400.2}
\shortauthors{Ergin et al.}

\begin{document}

\title{Recombining Plasma \& Gamma-ray Emission in the Mixed-morphology Supernova Remnant 3C 400.2}

\email{$^{\star}$ ergin.tulun@gmail.com, $^{\dagger}$aytap.sezer@avrasya.edu.tr, $^{\ddagger}$sano@a.phys.nagoya-u.ac.jp}
\author{T. Ergin$^{\star}$}
\affil{TUBITAK Space Technologies Research Institute, ODTU Campus, 06800, Ankara, Turkey}
\author{A. Sezer$^{\dagger}$}
\affil{Department of Electrical-Electronics Engineering, Faculty of Engineering and Architecture, Avrasya University, 61250 Trabzon, Turkey}
\author{H. Sano$^{\ddagger}$}
\affil{Department of Physics, Nagoya University, Chikusa-ku, Nagoya, Aichi 464-8601, Japan}
\author{R. Yamazaki}
\affil{Department of Physics and Mathematics, Aoyama Gakuin University, 5-10-1 Fuchinobe, Sagamihara 252-5258, Japan}
\and
\author{Y. Fukui}
\affil{Department of Physics, Nagoya University, Chikusa-ku, Nagoya, Aichi 464-8601, Japan}
\vspace{1cm}

\begin{abstract}
3C 400.2 belongs to the mixed morphology supernova remnant class, showing center-filled X-ray and shell-like radio morphology. We present a study of 3C 400.2 with archival {\it Suzaku} and {\it Fermi}-LAT observations. We find recombining plasma (RP) in the {\it Suzaku} spectra of north-east and south-east regions. The spectra of these regions are well described by two-component thermal plasma models: The hard component is in RP, while the soft component is in collisional ionization equilibrium (CIE) conditions. The RP has enhanced abundances indicating that the X-ray emission has an ejecta origin, while the CIE has solar abundances associated with the interstellar material. The X-ray spectra of north-west and south-west regions are best fitted by a two-component thermal plasma model: an ionizing and a CIE plasma. We have detected GeV gamma-ray emission from 3C 400.2 at the level of $\sim$5$\sigma$ assuming a point-like source model with a power-law (PL) type spectrum. We have also detected a new GeV source at the level of $\sim$13$\sigma$ assuming a Gaussian extension model with a PL type spectrum in the neighborhood of the SNR. We report the analysis results of 3C 400.2 and the new extended gamma-ray source and discuss the nature of gamma-ray emission of 3C 400.2 in the context of existing NANTEN CO data, DRAO H\,{\sc i} data, and the {\it Suzaku} X-ray analysis results. 
\end{abstract}


\section{Introduction}
3C 400.2 (also known as G53.6$-$2.2) is a mixed-morphology (MM) \citep{RhoPe98} Galactic supernova remnant (SNR) measured at a distance of 2.8 $\pm$ 0.8 kpc by the H\,{\sc i} observations \citep{Gi98}. The radio morphology of 3C 400.2 can be described as two overlapping circular shells of diameters 14$'$ and 22$'$ \citep{Du94} resulting from a single supernova (SN) explosion \citep{Sa95, Yo01, Sch06}, which is consistent with H\,{\sc i} observations showing a denser region to the north-west, where the smaller shell is located \citep{Gi98}. The optical morphology is characterized by a shell-like structure with a smaller radius of about 8$'$ \citep{Wi93, Am06}. H${\alpha}$ filaments correspond with regions of higher compression in the H\,{\sc i} contours and the center of the H\,{\sc i}  void coincides with the geometrical center of the optical shell \citep{Gi98}. 

{\it Einstein} Image Proportional Counter and {\it ROSAT} observations revealed that 3C 400.2 has centrally peaked X-ray emission and the X-ray features are not correlated with any radio features or optical filaments \citep{Sa95, Lo91}. The SNR was found to be $\sim$15000 yrs old \citep{Lo91}. {\it ASCA} observation showed that the X-ray spectra can be well fitted by thin thermal plasma models \citep {Yo01}. Using {\it Chandra} data, \citet {Br15} found recombining plasma (RP, overionization) in all parts of 3C 400.2. They found that the thermal X-ray emission of the whole remnant can be well explained by a two component non-equilibrium ionization (NEI) model: one component is underionized, has temperature $kT_{\rm e}$ $\sim$ 3.9 keV and super-solar abundances, the other component has temperature $kT_{\rm e}$ $\sim$ 0.14 keV, solar abundances and shows signs of RP. 
\begin{table*}
 \begin{minipage}{170mm}
  \begin{center}
  \caption{Log of the {\it Suzaku} XIS observations.}
   \vspace{0.2cm}
\begin{tabular}{@{}lccccc@{}}
  \hline
\hline
Name  &  Obs.ID  & Obs.Date & RA (deg) & Dec. (deg) & Exposure (ks)     \\
\hline
3C 400.2 NW &509068010 &2014-04-23&  294.5044 & 17.3912  &21.5  \\
3C 400.2 SW & 509069010 & 2014-04-14 & 294.5545 & 17.1470 & 24.2  \\
3C 400.2 SE &509070010 & 2014-04-23 & 294.8215  & 17.1213 &24.9  \\
3C 400.2 NE & 509071010 & 2014-04-23 & 294.7841  & 17.3445  & 20.2 \\
  \hline
\end{tabular}
\label{table_1}
\end{center}
\end{minipage}
\end{table*}

The first SNRs detected by the Large Area Telescope on board {\it Fermi} Gamma-Ray Space Telescope ({\it Fermi}-LAT) were mostly middle-aged MM SNRs interacting with molecular clouds (MCs), W51C, IC443, W28, W44, and W49B \citep{Ab09,Ab10a, Ab10b,Ab10c,Ab10d}. Interactions of these MM SNRs with MCs was shown by 1720 MHz OH masers \citep{Za95,Fr96,Gr97,Cl97,He09} and near-infrared observations \citep{Ke07}. 
Although 3C 400.2 is also a middle-aged MM SNR, no evidence of interactions of 3C 400.2 with MCs were reported and it was placed in the table of `SNRs Not Detected by the LAT' in the 1st SNR Catalog of {\it Fermi}-LAT \citep{Ac16}. Also in the TeV range of the gamma-ray emission, 3C 400.2 was not detected, where the upper limit on the flux at a 99\%  confidence level was reported to be 4.1$\times$10$^{-13}$ cm$^{-2}$ s$^{-1}$ by \citet{Bo11}. This value is close to the theoretical flux prediction of 3.6$\times$10$^{-13}$ cm$^{-2}$ s$^{-1}$ \citep{Bo11}. The SNR is located about 2$^{\circ}\!\!$.5 from the nearest gamma-ray emission region and about 1$^{\circ}$ from the pulsar PSR B1933$+$16. There are no other nearby counterparts to 3C 400.2. 

To date, RPs have been discovered in the {\it Suzaku} X-ray spectra of several MM SNRs (e.g., W49B: \citealt{Oz09}, IC 443: \citealt{Ya09}, 3C 391: \citealt{Er14}, and G166.0+4.3: \citealt{Ma17}). The RP is characterized with a higher ionization temperature than an electron temperature ($kT_{\rm z}$ $>$ $kT_{\rm e}$). Although the origin of RP in SNRs has not been understood yet, there is a strong correlation between the RPs and the MM SNRs. That the MM SNRs are generally associated with MC regions and GeV gamma-ray emission, indicates that they are mostly located in dense environments. Therefore, it is expected that they are originated in core-collapse supernovae (CC SNe). Thus, 3C 400.2 is one of the best candidates to search RP in SNRs.  A systematic study of RP SNRs will help us to understand the physical properties of the RPs in more detail. In the literature, RP has been mostly identified in the ejecta component of SNRs. On the other hand, for 3C 400.2  \citet {Br15} found RP in the interstellar medium (ISM) component. 

In this paper, we aim to investigate RP in different regions of 3C 400.2. For a detailed investigation,  we use archival data from {\it Suzaku}, which has high spectral resolution. We also investigate the GeV gamma-ray properties of 3C 400.2 using archival {\it Fermi}-LAT data. The observation and analysis methods are described in Section 2. In Sections 3 and 4, the detailed analysis results are reported and the nature of the X-ray and gamma-ray emission from 3C 400.2 is discussed.
\begin{figure}
\centering \vspace*{1pt}
\includegraphics[width=0.45\textwidth]{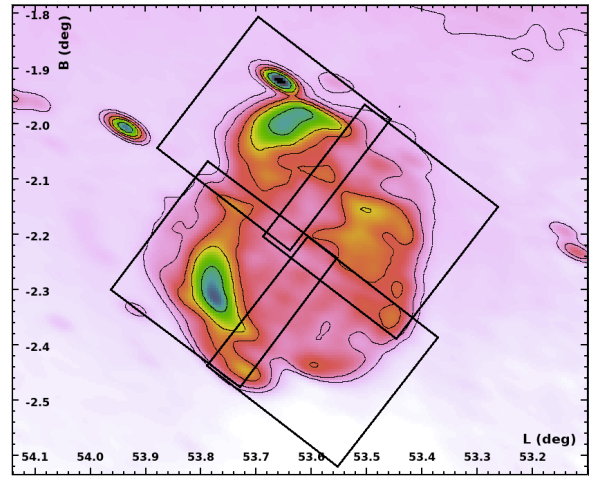}
\caption{The radio continuum data at 1420 MHz are from the Canadian Galactic Plane Survey (CGPS). The radio contours are at levels of 6.5, 7.0, 8.0, 9.0, and 11.0 mJy beam$^{-1}$. The FoV of XIS observations are shown by boxes.}
\label{figure_1}
\end{figure}
\begin{figure}
\centering \vspace*{1pt}
\includegraphics[width=0.45\textwidth]{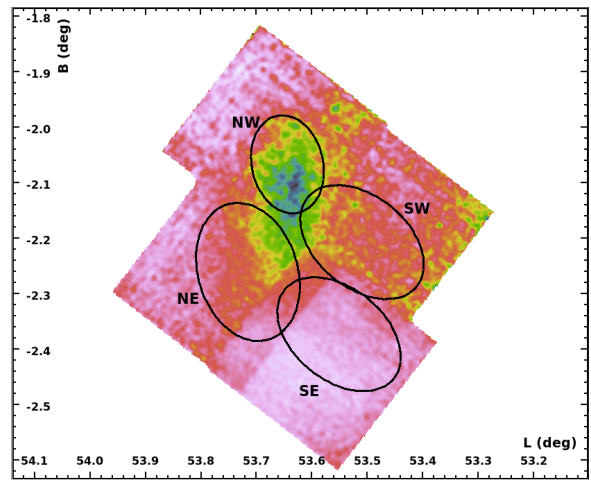}
\caption{Exposure-corrected Suzaku XIS mosaic image of 3C 400.2 in the 0.6-4.0 keV energy band. The black ellipses indicate the spectral extraction regions.}
\label{figure_1a}
\end{figure}

\section{Analysis}
\subsection{X-ray Observation and Data Reduction}
{\it Suzaku} observed the north-west (NW), south-east (SE), north-east (NE) and south-west (SW) regions of 3C 400.2 with X-ray Imaging Spectrometer \citep[XIS;][]{Ko07}. The details of these pointings are summarized in Table \ref{table_1}. The fields of views (FoVs) of these observations are shown in Figure \ref{figure_1} with the squares overlaid on the Canadian Galactic Plane Survey \citep[CGPS;][]{Ta03} radio image taken with Dominion Radio Astrophysical Observatory (DRAO). The XIS consists of four X-ray CCD cameras, each of which is located in the focal plane of the X-Ray Telescope \citep[XRT;][]{Se07}. The XIS0, 2 and 3 are front-illuminated (FI) CCDs, the other (XIS1) is a back-illuminated (BI) CCD. The XIS has a sensitivity in an energy range of 0.2$-$12.0 keV. The FoV of the XIS is $\sim$17\farcm8$\times$17\farcm8. The XIS2 and a small fraction of XIS0 have been dysfunctional since 2006 November 9\footnote{http://www.astro.isas.jaxa.jp/suzaku/news/2006/1123/} and 2009 June 23\footnote{http://www.astro.isas.jaxa.jp/suzaku/news/2009/0702/}, respectively, because of the damage by a possible micro-meteorite. We thus used the data obtained with the remaining part of XIS0, XIS1, and XIS3 CCD cameras. We downloaded {\it Suzaku} archival data of 3C 400.2 from the Data Archives and Transmission System (DARTS)\footnote{http://www.darts.isas.jaxa.jp/astro/suzaku/}.
The data reduction and analysis were carried out with {\sc headas} package version 6.20 and {\sc xspec} version 12.9.1 \citep {Ar96} with {\sc atomdb} 3.0.8  \citep {Sm01, Fo12}. For extracting images and spectra we used {\sc xselect} version 2.4d. We combined 3$\times$3 and 5$\times$5 event files using {\sc xis5x5to3x3} and {\sc xselect}. In the spectral analysis, the redistribution matrix files (RMFs) and the ancillary response files (ARFs) are generated by the {\sc xisrmfgen} and {\sc xissimarfgen} tools, respectively \citep {Is07}. 
\begin{figure*}
\centering \vspace*{1pt}
\includegraphics[width=0.45\textwidth]{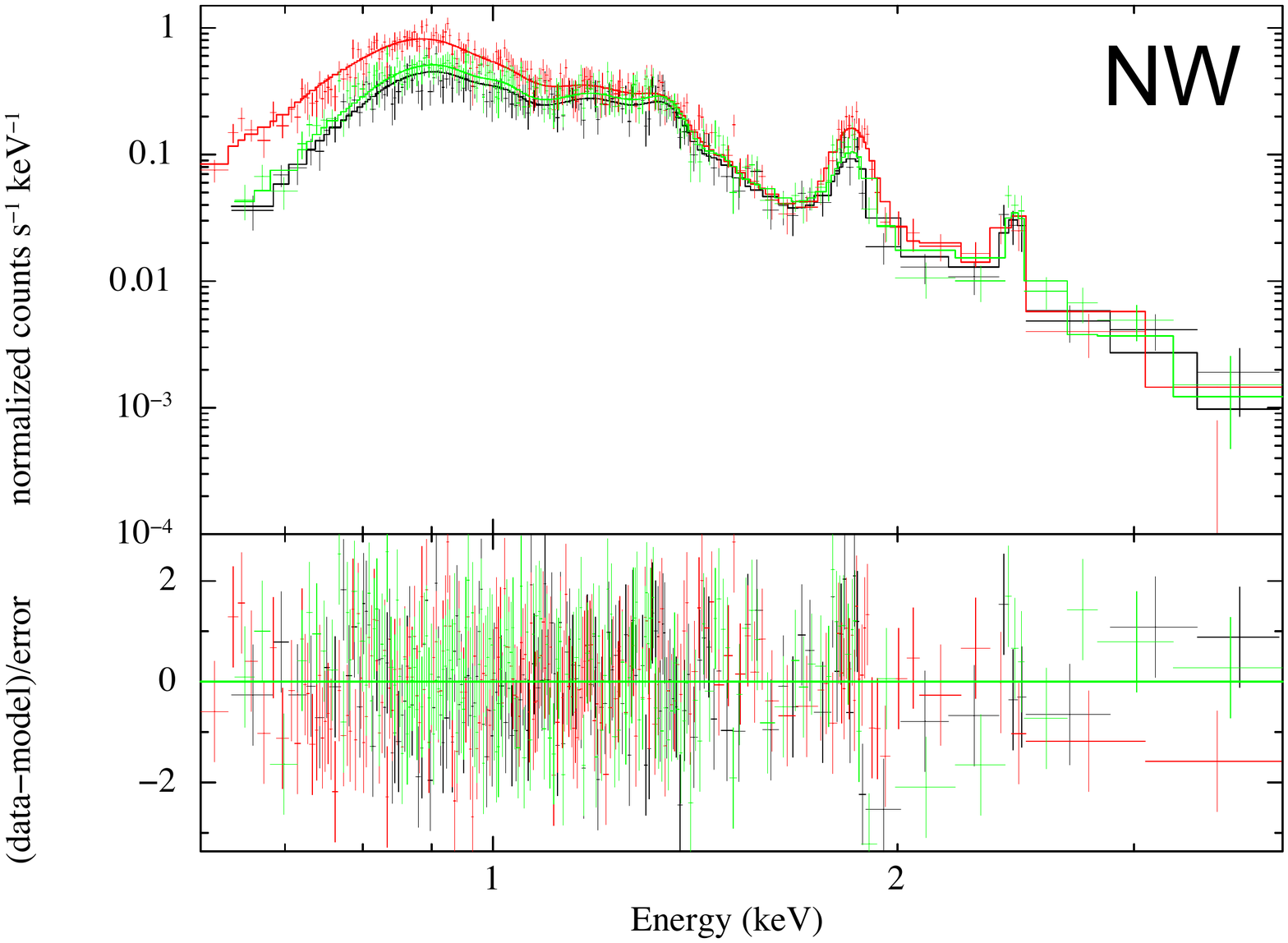}
\includegraphics[width=0.45\textwidth]{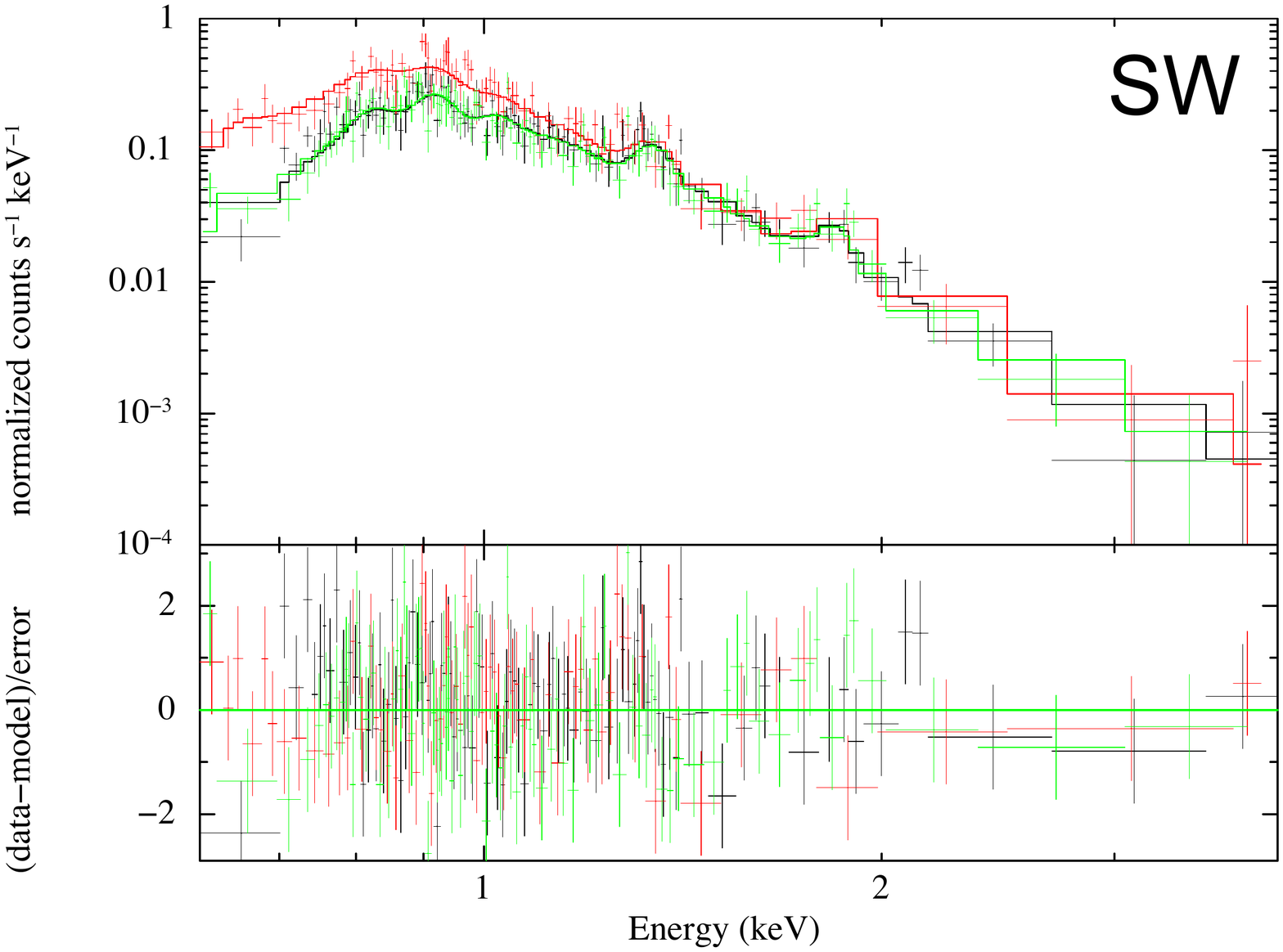}
\includegraphics[width=0.45\textwidth]{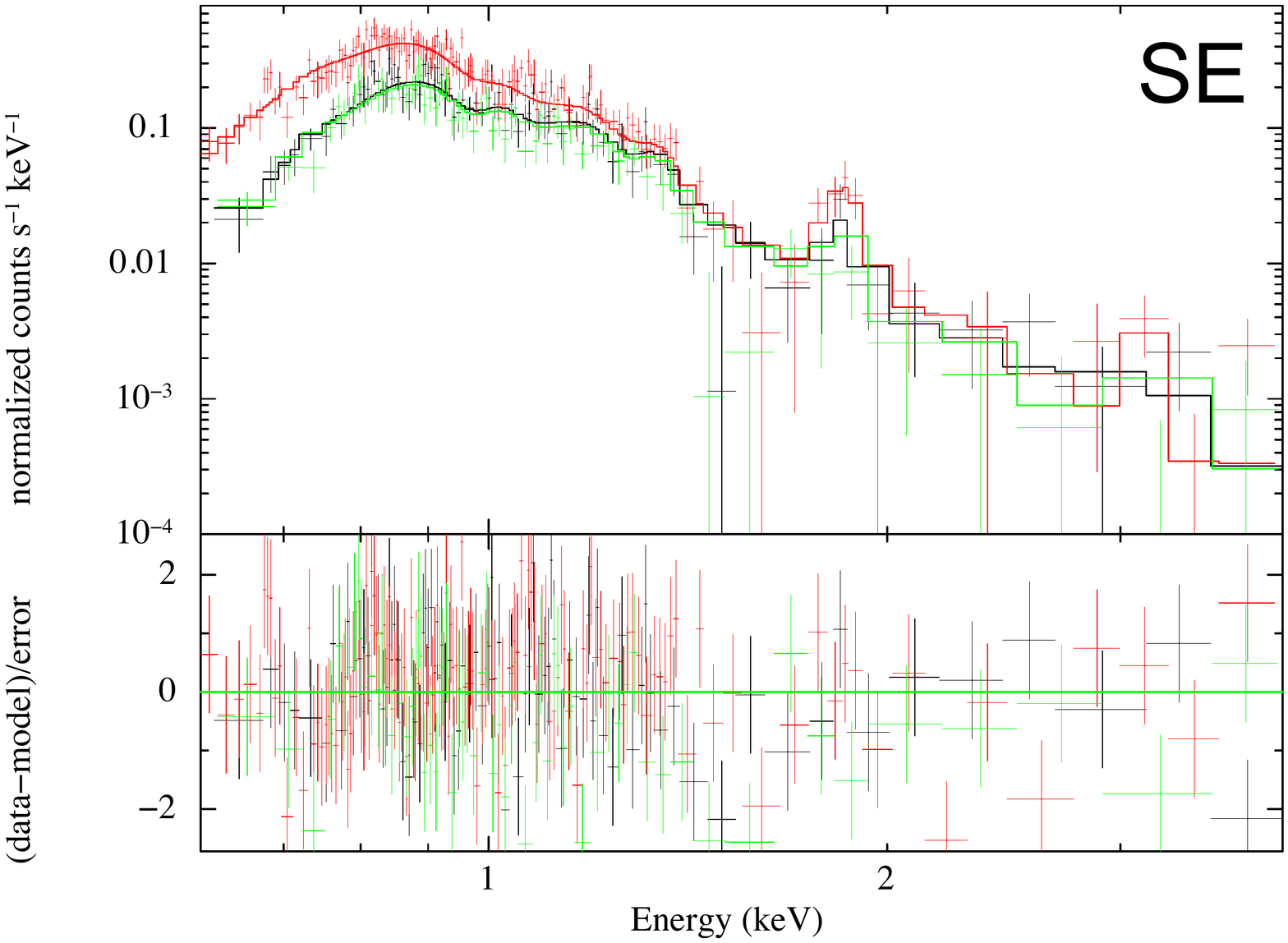}
\includegraphics[width=0.45\textwidth]{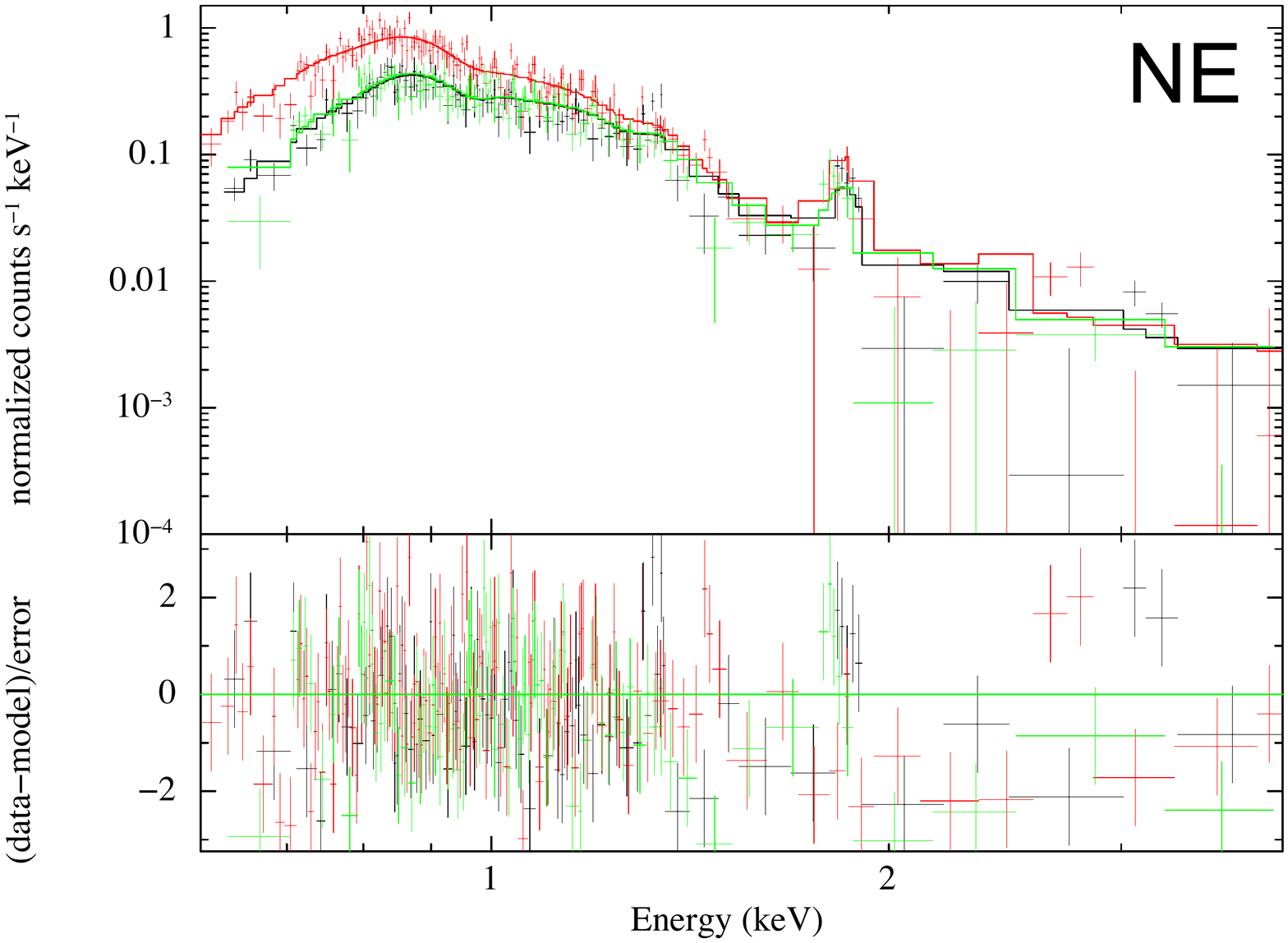}
\vspace{-0.5cm}
\caption{Background-subtracted XIS (XIS0:black, XIS1:red, XIS3:green) spectra of the NW, SW, SE, and NE regions of 3C 400.2 in the 0.6$-$4.0 keV energy band. The spectra are overlaid with the best-fit model. The bottom panels show the residuals of the best-fit.}
\label{figure_2}
\end{figure*}

\subsection{X-ray Spectral Analysis} \label{Spectral analysis}
\subsubsection{Background Subtraction}
In order to construct the background spectra, we extracted data from source free regions of the same observations. The calibration sources at the corners of the FoVs were excluded. The background consists of the instrumental Non-X-ray Background (NXB) and the X-ray Background for 3C 400.2. Similar to \citet {Ma17}, we adopted a model by \citet {Ma09} for the X-ray background modeling. This consists of the Cosmic X-ray background (CXB), the Local Hot Bubble (LHB), and two thermal components for the Galactic Halo (GH$_{\rm cold}$ and GH$_{\rm hot}$). Since the SNR is located at far from the Galactic Center, the Galactic ridge X-ray emission is negligible. The NXBs were estimated by {\sc xisnxbgen} \citep {Ta08} and subtracted from both the source and the background data. In the fits, we fixed the electron temperatures of the LHB, GH$_{\rm cold}$ and GH$_{\rm hot}$ at the values given by \citet {Ma09}. The absorption parameter and the normalization of each component were left as free parameters. The photon index of the CXB was fixed at 1.4 and the normalization was left as a free parameter. Then, we subtracted it from the source spectra. All spectra were grouped with a minimum of 25 counts bin$^{-1}$ using the FTOOL {\sc grppha}. Using the {\sc fakeit} command in {\sc xspec}, we simulated the background spectra. 

\vspace{0.5cm}
\subsubsection {Spectral Fitting}
Figure \ref{figure_1a} shows XIS mosaic image of 3C 400.2 in the 0.6-4.0 keV energy band. We analyzed spectra extracted from each region marked in Figure \ref{figure_1a} as ellipses. For spectral fitting, we used a non-equilibrium recombining collisional plasma model with variable abundances (VRNEI\footnote{heasarc.gsfc.nasa.gov/xanadu/xspec/manual/XSmodelRnei.html} model with NEI version 3.0 in {\sc xspec}). The VRNEI is based on the assumption that the plasma changes from the collisional equilibrium state, with the initial temperature $kT_{\rm init}$ to the non-equilibrium recombining state, with the temperature $kT_{\rm e}$. In our analysis the $kT_{\rm init}$ is a free parameter for the RP model, whereas it is fixed at 0.0808 keV for the IP model. For CIE plasma model, we used VRNEI model with ionization parameter ($\tau$=$n_{\rm e}t$) fixed at 1$\times10^{13}$ cm$^{-3}$ s. We employed TBABS model in {\sc xspec} \citep {Wi00} for interstellar absorption with the abundances set to \texttt{wilm} \citep {Wi00}.

In order to search for RP, we first applied a one-component RP model to the NE spectra with an absorption \citep[TBABS:][]{Wi00} model, TBABS$\times$VRNEI. In this fitting, the absorption ($N_{\rm H}$), electron temperature ($kT_{\rm e}$), initial temperature ($kT_{\rm init}$), normalization and the ionization parameter ($\tau$=$n_{\rm e}t$) were set as free parameters, where $n_{\rm e}$ and $t$ are the electron density and elapsed time following shock-heating. The abundances of Ne, Si, and S  were left as free parameters, while the other abundances were fixed to the solar abundance \citep {Wi00}. This model gave a large reduced-${\chi}^2$ value. Then we fitted the spectra with a two-component VRNEI plasma model, an RP and a CIE model. For the CIE component, $kT_{\rm e}$, $kT_{\rm init}$ and normalization parameters were set free, while the abundances of all elements were fixed to the solar values assuming that the emission is from the shocked ISM. The spectra are well reproduced by a two-component model with an acceptable reduced ${\chi}^2$ value. We found that the plasma of NE region is in the RP state. 

We next applied the same steps for the SE, SW and NW spectra. The spectral analyses showed that these regions require two thermal components, indicating that the plasma contains a hard and soft temperature component. The plasma in the SE region consists of the RP and CIE component. For high temperature component of SW and NW regions, we found that the initial temperature ($kT_{\rm init}$) is lower than electron temperature ($kT_{\rm e}$) indicating that the plasma is not in the recombining phase. The best-fit parameters of each region are summarized in Table \ref{table_2} with the corresponding errors at the 90\% confidence limit. The spectra of all the regions are presented in Figure \ref{figure_2}.
\begin{table*}
 \begin{minipage}{170mm}
  \caption{Best-fit X-ray spectral parameters of 3C 400.2.}
\renewcommand{\arraystretch}{1.5}
\centering
  \begin{tabular}{@{}p{1.7cm}p{4cm}p{2.1cm}p{2.1cm}p{2.1cm}p{2.1cm}@{}}
  \hline\hline
     Component& Parameters & NW & SW & NE & SE\\
\hline
Absorption & $N_{\rm H}$ ($\times10^{22}$ cm$^{-2})$ & $0.53_{-0.02}^{+0.03}$  & $0.39_{-0.01}^{+0.02}$ & $0.35_{-0.01}^{+0.02}$ & $0.32_{-0.02}^{+0.01}$ \\
VRNEI & $kT_{\rm e}$ (keV)&  $0.74_{-0.01}^{+0.01}$  &   $0.69_{-0.01}^{+0.02}$    &    $0.71_{-0.02}^{+0.03}$  & $0.63_{-0.01}^{+0.02}$ \\
& $kT_{\rm init}$ (keV) &    $0.0808$  (fixed)  &   $0.0808$  (fixed)   &   $3.15_{-0.63}^{+0.34}$   & $1.32_{-0.05}^{+0.07}$ \\
&   Ne  & $3.3_{-0.2}^{+0.4}$ &   $1.4_{-0.2}^{+0.5}$    &   $2.4_{-0.6}^{+0.2}$   & $2.1_{-0.1}^{+0.5}$ \\
&  Si  & $2.1_{-0.4}^{+0.3}$ &   $1.2_{-0.1}^{+0.1}$    &  $1.5_{-0.3}^{+0.4}$    & $1.6_{-0.3}^{+0.1}$  \\
&  S  & $2.3_{-0.1}^{+0.3}$  &  $1.8_{-0.2}^{+0.3}$     &    $1.9_{-0.3}^{+0.4}$  & $2.8_{-1.1}^{+0.5}$ \\
&   $\tau$ ($\times10^{11}$ cm$^{-3}$ s)&  $1.5_{-0.1}^{+0.2}$&    $1.7_{-0.2}^{+0.3}$   &   $1.8_{-0.5}^{+0.2}$   & $2.1_{-0.6}^{+0.4}$ \\
&  Norm ($\times10^{-4}$ ph cm$^{-2}$ s$^{-1}$)  &  $8.03_{-1.54}^{+1.45}$  &   $6.12_{-1.13}^{+1.26}$    & $4.43_{-1.14}^{+1.06}$     & $3.12_{-1.14}^{+0.27}$ \\
\hline
VRNEI & $kT_{\rm e}$ (keV)&  $0.34_{-0.01}^{+0.02}$  &   $0.41_{-0.03}^{+0.02}$    &    $0.47_{-0.02}^{+0.03}$  & $0.35_{-0.01}^{+0.02}$ \\
&   $\tau$ ($\times10^{13}$ cm$^{-3}$ s)&  $1$ (fixed) &    $1$ (fixed)   &   $1$ (fixed)   & $1$ (fixed) \\
&  Norm ($\times10^{-3}$ ph cm$^{-2}$ s$^{-1}$)  &  $3.45_{-0.24}^{+0.21}$  &   $2.58_{-0.61}^{+0.47}$    & $1.98_{-1.21}^{+0.17}$     & $3.11_{-0.98}^{+0.77}$ \\
\hline
& reduced-$\chi^{2}$ (dof) & 1.07 (793) &   1.02 (864)    &  1.05 (713)    & 1.08 (832) \\
 \hline
\end{tabular}
\label{table_2}
\end{minipage}
\end{table*}
 
\subsection {Gamma-ray Analysis} \label{gamma_analysis}
To search for a gamma-ray emission in the GeV energy range, we analyzed the gamma-ray data of {\it Fermi}-LAT for the time period of 2008-08-04 $-$ 2017-02-06. In this paper we used the Fermi analysis toolkit \texttt{fermipy} version 0.8.0\footnote{http://fermipy.readthedocs.io/en/latest/index.html}.   
Using \texttt{gtselect} of Fermi Science Tools (FST), we selected the {\it Fermi}-LAT Pass 8 `Source' class and front$+$back type events coming from zenith angles smaller than 90$^{\circ}$ and from a circular region of interest (ROI) with a radius of 30$^{\circ}$ centered at the SNR's radio position. The maximum likelihood fitting method \citep{Ma96} was employed on the spatially and spectrally binned data using the P8R2$_{-}$SOURCE$_{-}\!\!$V6 version of the instrument response function. After the maximum likelihood fitting between 200 MeV and 300 GeV, the detection significance value is calculated, which is roughly the square root of the test statistics (TS) value and larger TS values indicate that the null hypothesis (maximum likelihood value for a model without an additional source) is incorrect.

The model of the analysis region contains the diffuse background sources and all the point-like and extended sources from the 3rd {\it Fermi}-LAT Source Catalog \citep{Ac15} located within a square region with a side of 15$^{\circ}$ centered on the ROI center. The normalization parameters of sources that are within 3$^{\circ}$ are set free. All parameters of the diffuse Galactic emission (\emph{gll$_{-}$iem$_{-}$v6.fits}) and the isotropic component (\emph{iso$_{-}$P8R2$_{-}$SOURCE$_{-}\!\!$V6$_{-}\!$v06.txt}) were freed. We also freed the normalization parameter of all sources with TS $>$ 10 and fixed all the parameters for sources with TS $<$ 10. 

The initial TS map was produced for a 10$^{\circ}$ $\times$ 10$^{\circ}$ analysis region using this model showing gamma-ray excess close to the location of 3C 400.2 and at other locations within the analysis region. In order to obtain the exact TS value and location of the excess regions, we used an iterative source-finding algorithm in \texttt{fermipy}, called \texttt{find$_{-}$sources}, which takes peak detection on a TS map to find new source candidates. The algorithm identified peaks with a significance threshold value higher than 3$\sigma$ and taking an angular distance of at least 1$\!^{\circ}\!\!$.5 from a higher amplitude peak in the map. It orders the peaks by their TS values and adds a source at each peak starting from the highest TS peak. Then it sets the source position by fitting a 2D parabola to the log-likelihood surface around the peak maximum. After adding each source, having a significance value above 5$\sigma$, it re-fits the spectral parameters of that source. With this algorithm we identified 9 new sources within the analysis region. The algorithm also listed the sources with a significance between 3$\sigma$ and 5$\sigma$. One of these sources, PS J1938.6$+$1722 was within the 95\% confidence radius of the 3C 400.2 position. So, throughout this paper we use PS J1938.6$+$1722 for 3C 400.2. 

As a next step we analyzed the closer vicinity of 3C 400.2 by taking an analysis region of 5$^{\circ}$ $\times$ 5$^{\circ}$. We followed the same analysis procedure as described for the 10$^{\circ}$ $\times$ 10$^{\circ}$ analysis region, except that we added two sources into the background model, which are PS J1934.5$+$1845, being the only bright gamma-ray source having a significance above 5$\sigma$ within the analysis region and PS J1938.6$+$1722 representing 3C 400.2 itself. 

Except finding the best-fit location and extension model, as well as producing the spectrum of PS J1934.5$+$1845, we did not go further into analyzing the time variability of the gamma-ray emission, because this source is about 1$\!^{\circ}\!\!$.8 away from 3C 400.2 and we wanted to keep this paper focused on the gamma-ray emission from 3C 400.2. After finding the best-fit locations of these two sources using the \texttt{localize} algorithm and the best-fit extension model for PS J1934.5$+$1845, we computed the TS map including PS J1934.5$+$1845 and PS J1938.6$+$1722 in the background model and recomputed it removing these two sources from the model. All results are given in Section \ref{subsection:Gamma-rayResults}. 

\subsection {CO, H\,{\sc i} and Radio Continuum Datasets}
To understand the atomic and molecular environment around 3C 400.2, we took $^{12}$CO (J = 1$-$0) data at 2.6 mm from the NANTEN Galactic Plane Survey \citep[NGPS;][]{Mi04}. The survey was made with the NANTEN 4-m radio telescope at the Las Campanas Observatory in Chile during the period from 1999 to 2003. The angular resolution of the telescope was 2\farcm6 in 115 GHz. The survey grid spacing was 4$'$ for the 3C 400.2 region. The velocity resolution and typical rms-noise fluctuations were 1.0 km s$^{-1}$ per channel and $\sim$0.25 K, respectively.

The H\,{\sc i}  data at 21 cm and the radio continuum data at 1420 MHz are from CGPS \citep{Ta03} carried out at DRAO. The angular resolution of datasets was $\sim$1$'$. The velocity resolution of H\,{\sc i}  datasets was $\sim$0.82 km s$^{-1}$ per channel. The typical rms-noise fluctuations of H\,{\sc i}  and continuum were $\sim$3 K per channel and $\sim$0.3 mJy beam$^{-1}$, respectively.

\section{Results and Discussion}
\begin{figure*}
\centering 
\includegraphics[width=0.8\textwidth]{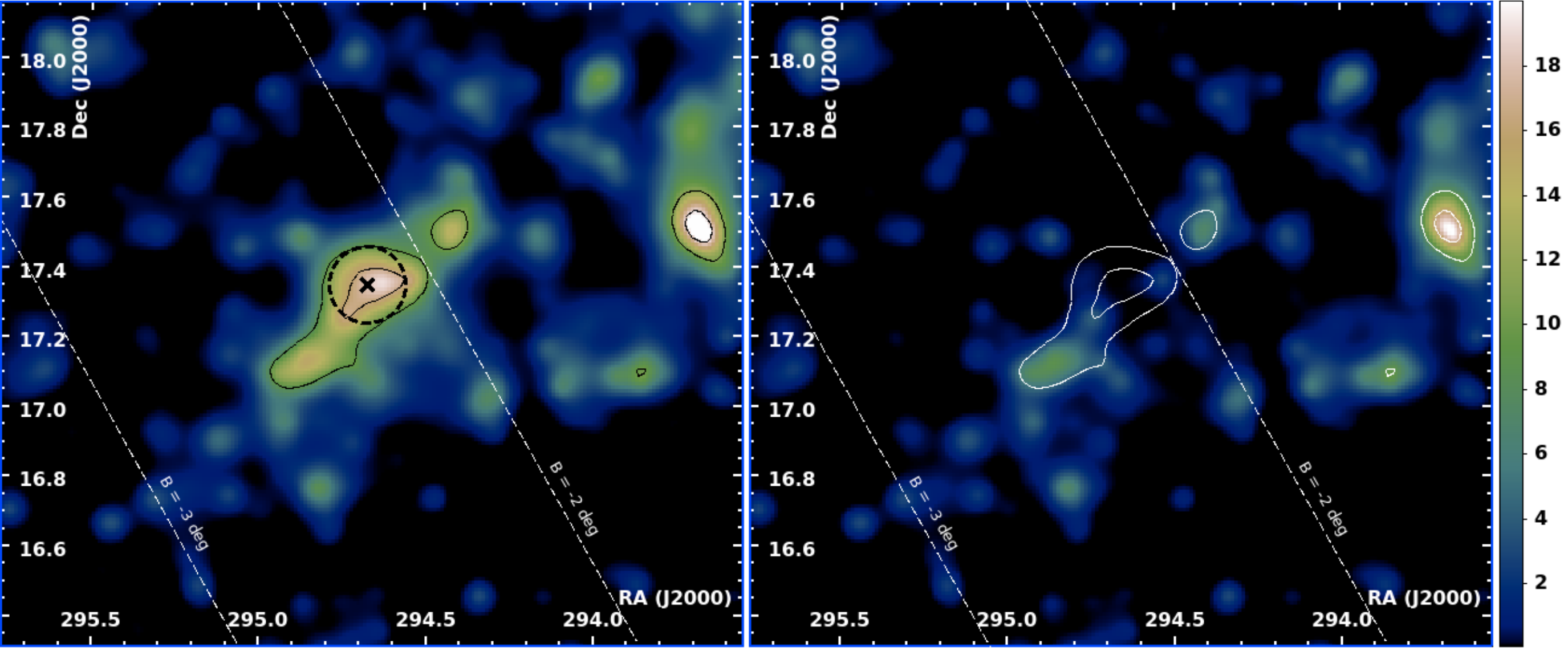}
\caption{Left Panel shows the gamma-ray TS map produced by excluding 3C 400.2 from the background model. The black contours represent the TS values at levels of 9 and 16. The black cross and black dashed circle are used for the best-fit location of 3C 400.2 and error circle at the 95\% confidence level, respectively. Right Panel is the TS map produced including 3C 400.2 in the background model. The color scale is set between 0.0 and 20.0 for both panels. The white dashed lines correspond to the Galactic latitudes of $-$2$^{\circ}$ and $-$3$^{\circ}$. }
\label{figure_3}
\end{figure*}

\subsection{X-ray Spectral Characteristics of Each Region}
In our spectral analysis, we searched for RPs and investigated X-ray spectral characteristics of each region in 3C 400.2. We found RPs in the NE and SE regions of 3C 400.2. The X-ray spectra of NE and SE regions consist of two thermal components: the hard component is in RP, while the soft component is in CIE. The hot plasma shows enhanced abundances of Ne, Si, and S indicating the RP has an ejecta origin. The X-ray emission of SW and NW regions is well represented by a two-component thermal plasma model; a high-temperature in NEI and a low-temperature in CIE condition. 

\citet {Br15} analyzed {\it Chandra} data and found RP in the ISM component (outer region) and under-ionized plasma in the ejecta component (inner region) in all regions of the remnant. Unlike this result, our analyses show RP in the ejecta component. We found small variations in the absorbing column density ($N_{\rm H}$), which is higher in the NW region than in other regions. We note that the absorption we obtained was lower than that of \citet {Br15}. An increasing $N_{\rm H}$ towards the NW region of the SNR indicates that the remnant is expanding into dense ISM in the NW region. For the hard component, the electron temperatures were obtained to be $\sim$0.71, $\sim$0.74, $\sim$0.63, and $\sim$0.69 keV for NE, NW, SE, and SW regions, respectively. The $kT_{\rm e}$ values show no significant difference between the regions. {\it ASCA} results show that the electron temperatures of all the regions range between $\sim$0.54 and $\sim$0.83 keV \citep{Yo01}, which are consistent with the electron temperatures found in our analysis. The under-ionized plasma temperatures found by \citet{Br15} are $\sim$3.23 and $\sim$3.59 keV for region 4 and region 2, respectively. These values are significantly higher than the electron temperatures found in our analysis. 

The inconsistencies between the results of \citet {Br15} and our results may be due to the differences of the procedures in the spectral analysis, which are listed here: (i) In order to obtain more reliable spectra, we used a model by \citet {Ma09} for the Galactic background in our background analysis. (ii) They used NEI model in {\sc spex} version 2.03 \citep{Ka96} for the spectral modeling, while we used VRNEI model in {\sc xspec} version 12.9.1 \citep{Ar96}. (iii) In our analysis we used the abundance table by \citep{Wi00} and the newly updated ATOMDB 3.0.8, while \citet{Br15} use the table derived by \citet{An89}.

Using the emission measure ($EM=n_{\rm e}n_{\rm H}V$, where $V$ is the volume of the X-ray emitting plasma) and assuming a distance of 2.8 kpc, we estimated the electron density $n_{\rm e}$ as $\sim$0.58$f^{-1/2}$, 0.51$f^{-1/2}$, 0.43$f^{-1/2}$, and 0.63 $f^{-1/2}$cm$^{-3}$ (where $f$ is the filling factor) for NW, NE, SE, and SW regions, respectively.

\subsection{Origin of the Recombining Plasma}
Several scenarios are proposed to explain the formation of RPs in SNRs. The RP is one signature of rapid electron cooling in SNRs. One is rapid electron cooling by the thermal conduction (e.g., \citealt{Co99, Sh99}) or adiabatic expansion (rarefaction) (e.g., \citealt{It89, Sh12}). In the following, we discuss the origin of the observed RP in 3C 400.2:

(i) We first discuss the thermal conduction scenario, where the hot ejecta in the SNR interior may cool by exchanging heat efficiently with the exterior material. To evaluate the possibility of the thermal condition scenario, we estimated the thermal conduction time-scale ($t_{\rm cond}$) for 3C 400.2. If the electron cooling is due to thermal conduction, a time-scale of the thermal conduction derived from the Spitzer thermal conductivity \citep {Sp62, Zh14} can be written as $t_{\rm cond}$ $\approx$ $k n_{\rm e} l_{\rm T}^{2} / \kappa$ $\sim$ $56(n_{\rm e}/1\,cm^{-3})(l_{\rm T}/10\,{\rm pc})(kT/0.6\ {\rm keV})^{-5/2} ({\rm ln} \Lambda/32)$ kyr,  where $k$ is Boltzmann's constant, $n_{\rm e}$ is the electron density, $l_{\rm T}$ is the scale length of the temperature gradient, $\kappa$ is the collisional conductivity and ${\rm ln}\Lambda$ is the Coulomb logarithm. We estimated thermal conduction time-scale and found that the time-scale is much longer than the estimated age this remnant. Therefore, we concluded that the thermal conduction model is unlikely to be able to explain the RP in 3C 400.2.

(ii) Another possible scenario for the electron cooling is adiabatic expansion, where the SN blast wave expands through a dense circumstellar medium into a rarefied ISM.
For the investigation the possibility of this scenario, we estimated the recombining time-scale ($t_{\rm rec}$) using the best-fit ionization time-scale from Table \ref{table_2} and the electron density for NE and SE regions. From $t=\tau$/$n_{\rm e}$, we obtained the recombining time-scales of $\sim$11.2 $f^{1/2}$ and $\sim$15.5 $f^{1/2}$ kyr for NE and SE regions, respectively, which are comparable to the SNR age.
\begin{figure*}
\centering \vspace*{1pt}
\includegraphics[width=0.9\textwidth]{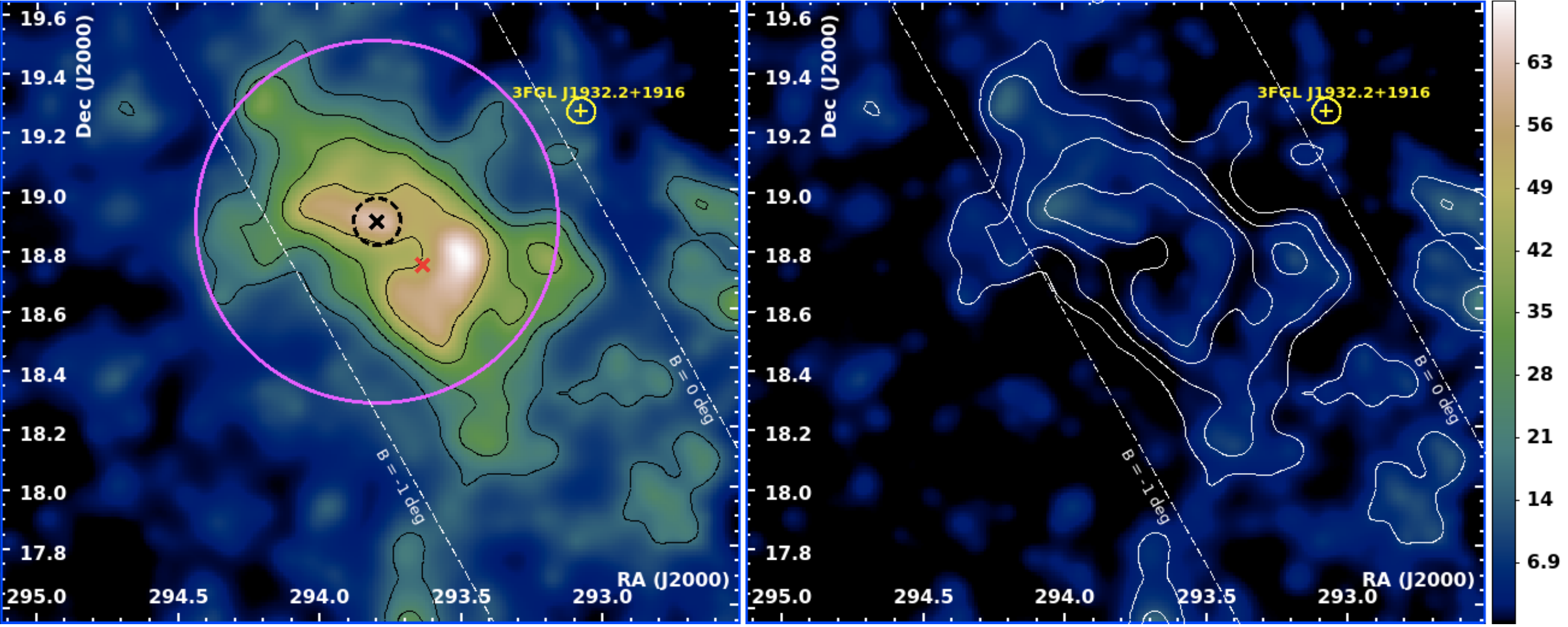}
\caption{TS map of the gamma-ray emission, where PS J1934.5$+$1845 was not included in the background model (Left Panel). The initial spatial model of the source was taken as point-like, where its location was chosen as the red cross shown in the Left Panel. The best-fit position of the point-like source is shown with a black cross and its positional error at a 95\% confidence level is shown as a black dashed circle. The magenta solid circle represents the best-fit extension found using a Radial Gaussian model centered around the best-fit position of PS J1934.5$+$1845. Right Panel shows the gamma-ray TS map, where PS J1934.5$+$1845 was included as a point-like gamma-ray source in the background model. The black and white contours on both Panels are the levels of the TS values, which are 16, 25, 36, 49, 69 and a gamma-ray source from the 3rd {\it Fermi}-LAT Source Catalog \citep{Ac15} is shown in yellow together with its positional error circle. The white dashed lines correspond to the Galactic latitudes of $-$1$^{\circ}$ and 0$^{\circ}$.} 
\label{figure_3b}
\end{figure*}
\begin{figure}
\centering 
\includegraphics[width=0.49\textwidth]{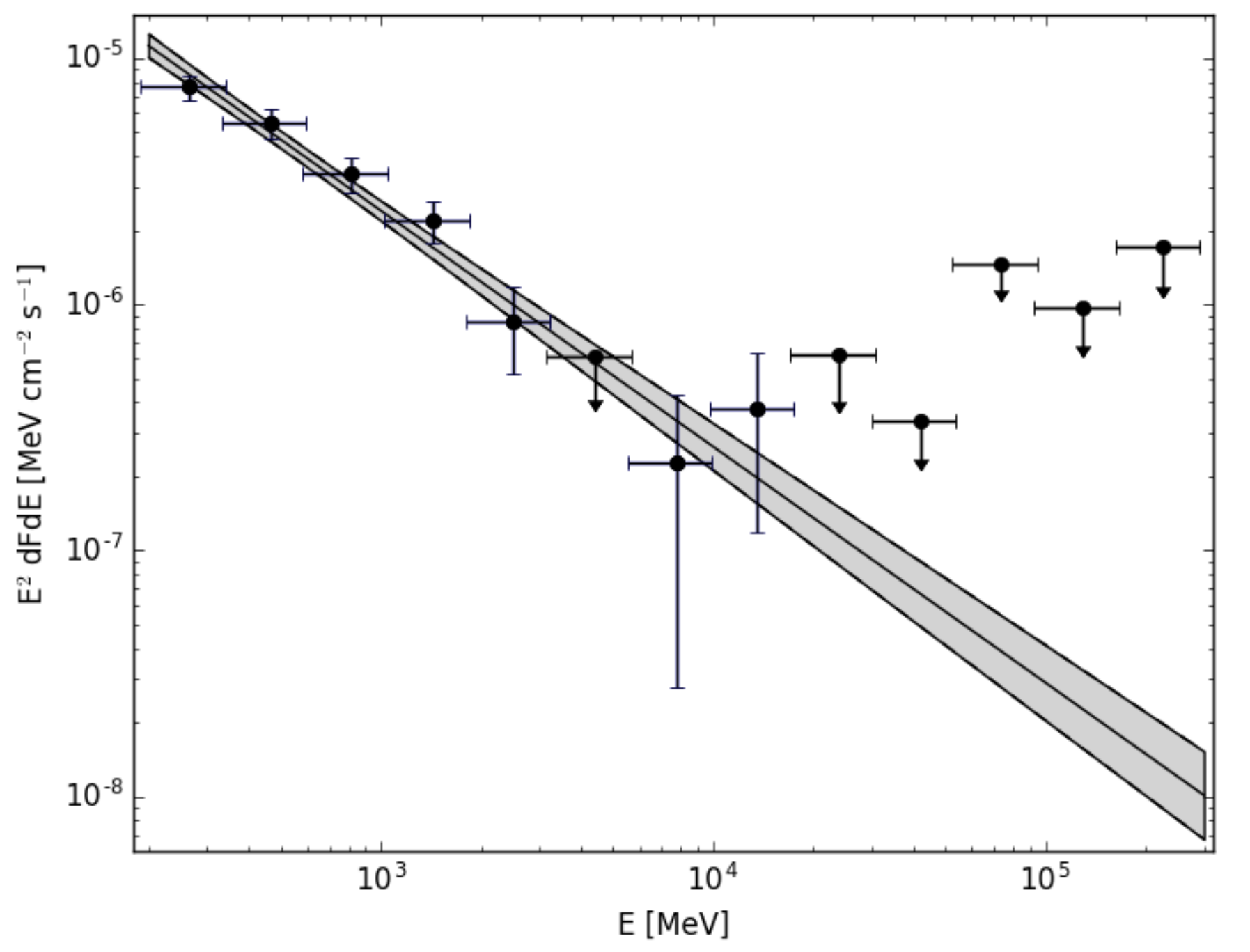}
\vspace{-0.5cm}
\caption{Gamma-ray spectral energy distribution of PS J1934.5$+$1845 assuming the SNR is a point-like source. The shaded region represents the model flux and its statistical errors obtained from fitting a PL-type spectrum to the given spectral data.}
\label{figure_3c}
\end{figure}

\subsection{Gamma-ray Results} \label{subsection:Gamma-rayResults}
\subsubsection{PS J1938.6$+$1722 (3C 400.2)}  
We detected a 5$\sigma$ gamma-ray excess from the direction of PS J1938.6$+$1722 assuming it as a point-like gamma-ray source, where the best-fit position was calculated to be R.A.(J2000) = 294$^{\circ}\!\!$.67 $\pm$ 0$^{\circ}\!\!$.11$^{\rm stat}$ and decl.(J2000) = 17$^{\circ}\!\!$.35 $\pm$ 0$^{\circ}\!\!$.11$^{\rm stat}$ ($\rm{R.A.~}(J2000) = 19^{\rm{h}} 38^{\rm{m}} 41\fs04$, $\rm{decl.~}(J2000)=17\degr 20\arcmin 56\farcs40$) using the \texttt{localize} method of \texttt{fermipy}. The spectrum was fit to power-law (PL), where the spectral index is found to be $\Gamma$ = 2.54 $\pm$ 0.24. The energy flux was found to be (2.77 $\pm$ 0.69) $\times$ 10$^{-6}$ MeV cm$^{-2}$ s$^{-1}$. Although the gamma-ray signal of 3C 400.2 is very weak, the TS map of the region centered around 3C 400.2's location shows an irregular morphology (Figure \ref{figure_3}), where the gamma-ray contours in black color represent the TS levels of 9 and 16. The Left Panel of Figure \ref{figure_3} shows the TS map, where 3C 400.2 is not included in the background model. After adding 3C 400.2 as a point-like source into the background model, most of the excess gamma rays at the location of 3C 400.2 disappeared, which is shown in the Right Panel of Figure \ref{figure_3}. Although in this analysis we fit the source by a point-like model, there are some ear-like extensions that are protruding out of the main bulk of the fitted gamma-ray source. When more statistics will be available in the future, the significance of these extensions may be better estimated.   

\subsubsection{PS J1934.5$+$1845} 
A new gamma-ray source, PS J1934.5$+$1845, was detected at a location of 1$^{\circ}\!\!$.83 away from the radio location of 3C 400.2. The detection significance was found to be $\sim$13$\sigma$ (i.e.$\!$ TS=161) assuming PS J1934.5$+$1845 as a point-like source during the source search procedure described in Section \ref{gamma_analysis}. Using the \texttt{localize} method of \texttt{fermipy}, the best-fit position for PS J1934.5$+$1845 was found to be R.A.(J2000) = 293$^{\circ}\!\!$.79 $\pm$ 0$^{\circ}\!\!$.08$^{\rm stat}$ and decl.(J2000) = 18$^{\circ}\!\!$.90 $\pm$ 0$^{\circ}\!\!$.08$^{\rm stat}$ ($\rm{R.A.~}(J2000) = 19^{\rm{h}} 35^{\rm{m}} 10\fs32$, $\rm{decl.~}(J2000)=18\degr 54\arcmin 07\farcs20$). 

The spectrum was fit to PL, where the spectral index is found to be $\Gamma$ = 2.98 $\pm$ 0.09. The total photon flux and energy flux was found to be (2.99 $\pm$ 0.31) $\times$ 10$^{-9}$ photons cm$^{-2}$ s$^{-1}$ and (1.20 $\pm$ 0.11) $\times$ 10$^{-6}$ MeV cm$^{-2}$ s$^{-1}$, respectively for the point-like source model having a PL-type spectrum. 

We used two models to parametrize the extended gamma-ray emission morphology of PS J1934.5$+$1845: Disk and Radial Gaussian models, where the width and location of the centers of the models were calculated by the \texttt{extension} method of \texttt{fermipy}. To detect the extension of a source, we used the TS of the extension (TS$_{\rm ext}$) parameter, which is the likelihood ratio comparing the likelihood for being a point-like source (L$_{\rm pt}$) to a likelihood for an existing extension (L$_{\rm ext}$), TS$_{\rm ext}$ = 2log(L$_{\rm ext}$/L$_{\rm pt}$). The `Extension Width', which is the 68\% containment radius of the extension model (R$_{68}$), was found to be 0$^{\circ}\!\!$.6107 $+$ 0$^{\circ}\!\!$.0866 $-$ 0$^{\circ}\!\!$.1141 with a TS$_{\rm ext}$ value of $\sim$40 for the Radial Gaussian model. As an extended source, PS J1934.5$+$1845 was detected with a significance of $\sim$13$\sigma$ (i.e.$\!$ TS=168). Assuming a PL-type spectrum for this extended source, we obtained $\Gamma$ = 2.38 $\pm$ 0.07 and the total photon flux and energy flux of PS J1934.5$+$1845 was found to be (2.54 $\pm$ 0.23) $\times$ 10$^{-8}$ photons cm$^{-2}$ s$^{-1}$ and (1.74 $\pm$ 0.16) $\times$ 10$^{-5}$ MeV cm$^{-2}$ s$^{-1}$, respectively. 
The Left Panel of Figure \ref{figure_3b} shows the different locations used for PS J1934.5$+$1845 during the analysis of the source: The initial fit location of the point-like PS J1934.5$+$1845 is shown with a red cross. The best-fit position for the point-like PS J1934.5$+$1845 is shown as a black cross and the statistical error of this position at a 95\% confidence level is shown as a black dashed circle. The magenta solid circle represents the best-fit extension of PS J1934.5$+$1845 with the center of it at the best-fit position, respectively. The spectral energy distribution of PS J1934.5+1845 as a point-like source having a PL-type spectrum is shown in Figure \ref{figure_3c}. The extended source spectrum is showing a similar shape. 
\begin{figure}
\centering
\includegraphics[width=0.48\textwidth]{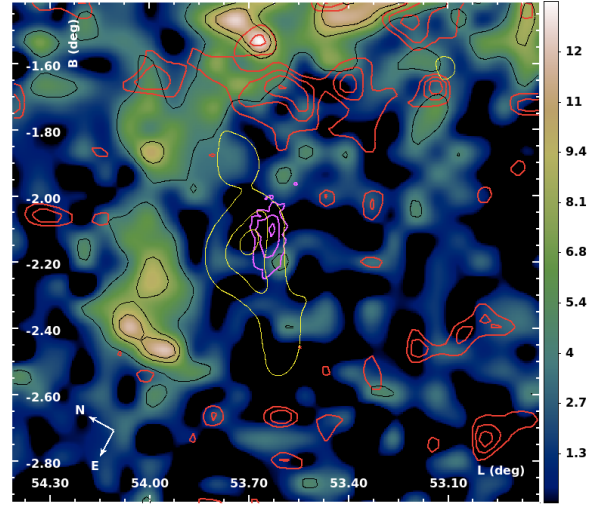}
\caption{CO intensity map is produced at the velocity range of 5$-$25 km s$^{-1}$ and the black contours show 4, 8, 10, and 13.5 K km s$^{-1}$.  The red contours overlaid represents the CO intensity at 2, 3, 4, and 5 K km s$^{-1}$ for the velocity range of 32$-$45 km s$^{-1}$. Yellow contours show the TS values for the gamma-ray emission at the levels of 9, 16, and 19 and magenta contours are for X-rays at 0.07, 0.09, 0.12, and 0.15 exposure-corrected counts.} 
\label{gfigure_4}
\end{figure}

\begin{figure}
\centering 
\includegraphics[width=0.5\textwidth]{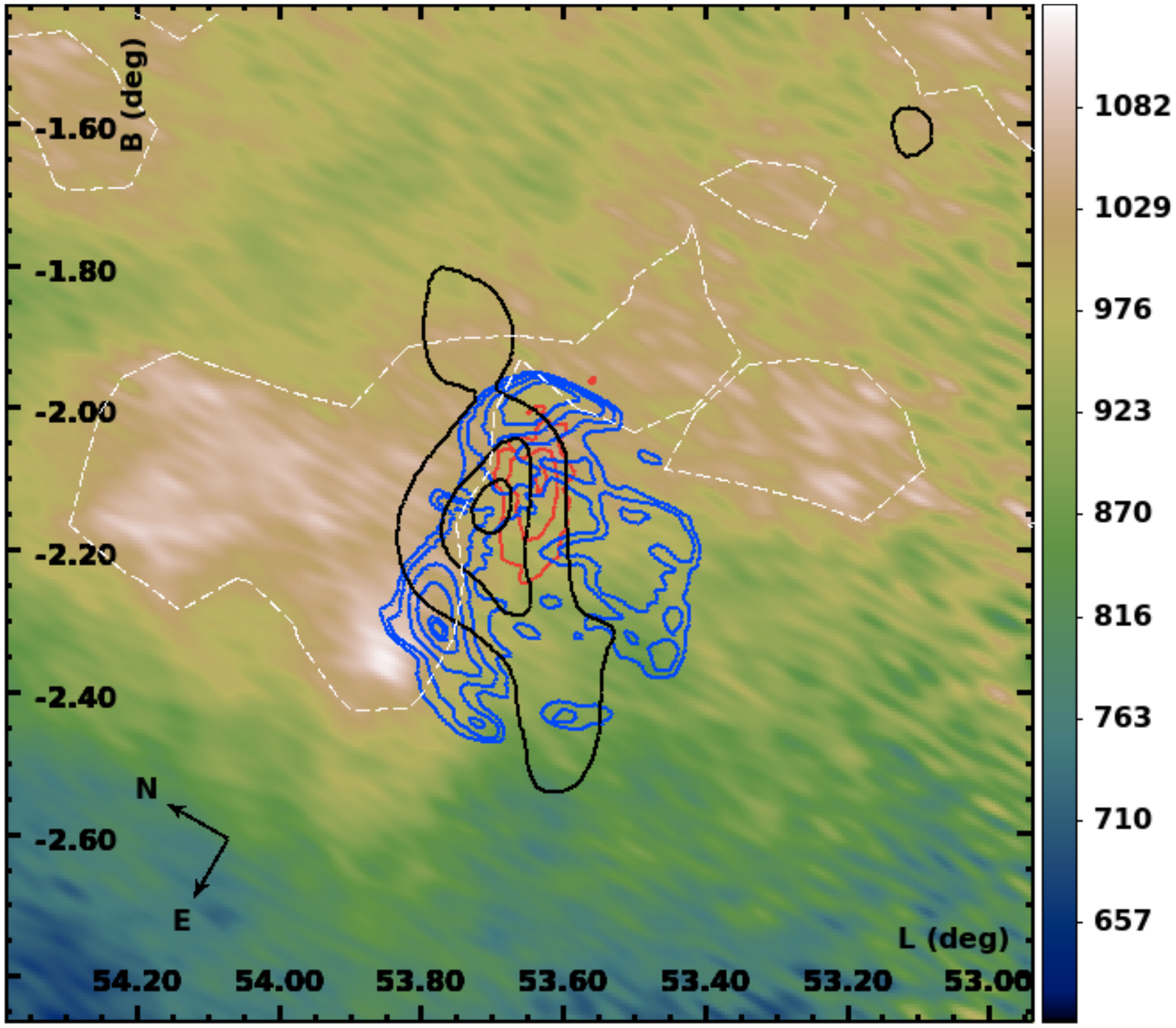}
\includegraphics[width=0.5\textwidth]{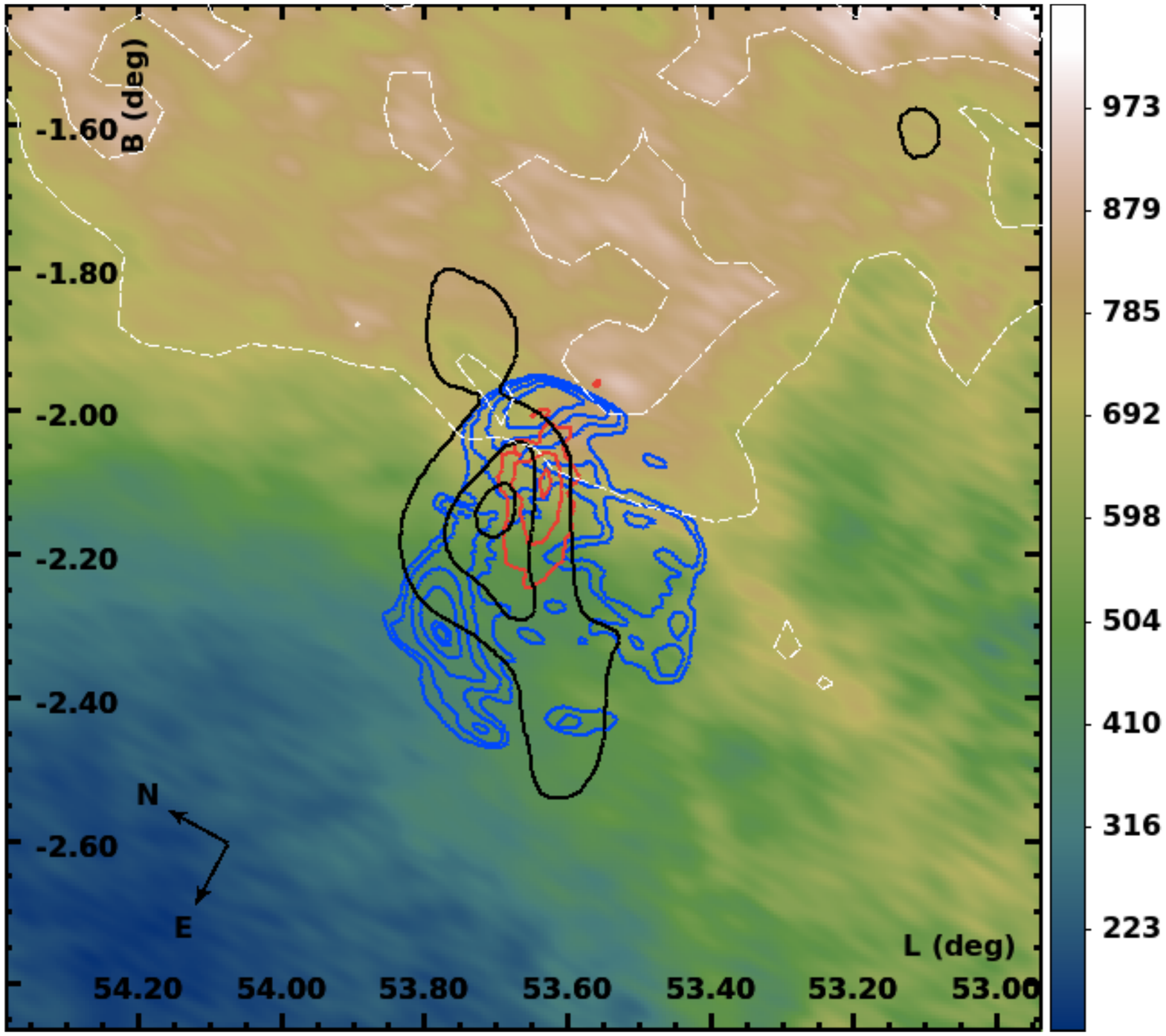}
\caption{ H\,{\sc i}  intensity maps produced at velocity ranges of 5$-$25 km s$^{-1}$ (Upper Panel) and 32$-$45 km s$^{-1}$ (Bottom Panel). The black contours represent the gamma-ray TS values of 3C 400.2  for 9, 11, and 16 and red contours are the X-ray counts at 0.07, 0.09, 0.12, and 0.15 exposure-corrected counts. The blue contours are for the radio continuum data as taken from DRAO at 7.5, 8, 9, 10, 11, 12.5, and 13 mJy beam$^{-1}$. The H\,{\sc i} intensity contours are shown in white dashed lines at 1010 and 1140 K km s$^{-1}$ for the velocity range of 5$-$25 km s$^{-1}$ (Upper Panel) and 129, 597, and 1066 K km s$^{-1}$ for the velocity range of 32$-$45 km s$^{-1}$ (Bottom Panel).}
\label{gfigure_5}
\end{figure}

\begin{figure}
\centering 
\includegraphics[width=0.45\textwidth]{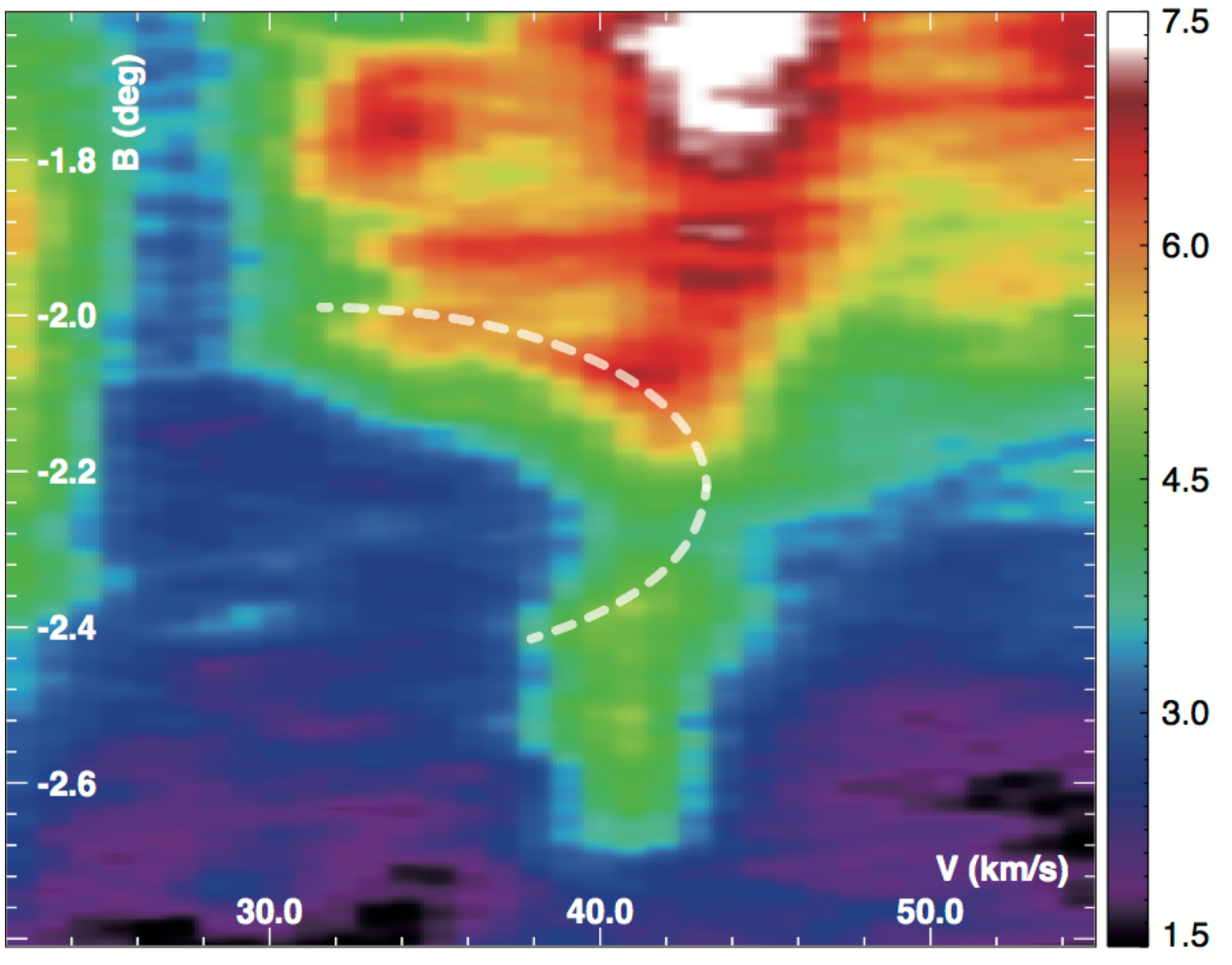}
\caption{Position-velocity diagram of H\,{\sc i} toward 3C 400.2. The integration range in the Galactic longitude is from 53$^{\circ}\!\!$.57 to 53$^{\circ}\!\!$.67. The dashed line represents the expanding shell of H\,{\sc i}.}
\label{gfigure_6}
\end{figure}

\subsection {The Environment of Molecular and Atomic Gases}
The interstellar gas plays an essential role in producing non-thermal radiation from the SNR. In particular, the interstellar protons in both the molecular and atomic clouds serve as targets for cosmic-ray (CR) protons to produce gamma rays via the hadronic process, which is the production mechanism of two gamma rays from the decay of a neutral pion created in a proton-proton interaction during the passage of SNR shocks through the dense MC. When the hadronic process is dominating the gamma-ray emission, we expect that the distribution of gamma rays shows a good spatial correspondence with that of the interstellar protons \citep{Fu12,Yo13,Fu14}. There are two hadronic scenarios for SNRs: The `crushed cloud' and `illuminated cloud' scenarios. In the former scenario, 
the hadronic gamma rays are a product of interactions between the MCs and the expanding shell of the SNR \citep{Bl82,Uc10}, while in the later scenario, relativistic protons escaping the SNR reach a nearby unshocked MC and produce hadronic gamma rays \citep{Ya06,Ga09,Oh11}, for which case there must be GeV/TeV sources found just outside the radio shell of the SNR. 

Figure \ref{gfigure_4} shows the CO intensity map toward 3C 400.2. We found two molecular clouds at the velocity range of 5$-$25 km s$^{-1}$ and 32$-$45 km s$^{-1}$. The former is located from the east (E) to the NW of the SNR with an incomplete shell-like structure, while the latter shows slightly weak emission in the western (W) part of the SNR. There are no dense molecular clouds spatially corresponding to the gamma-ray emission.

Figure \ref{gfigure_5} shows the H\,{\sc i}  intensity maps in the velocity range of 5$-$25 km s$^{-1}$ and 32$-$45 km s$^{-1}$. We noted that the 5$-$25 km s$^{-1}$ cloud is distributed in the NE and the NW, which form the cavity-like H\,{\sc i}  structure along the radio continuum shell. Further, the intensity peak of radio continuum in the NE (l, b $\sim$ 53$^{\circ}\!\!$.8, $-$2$^{\circ}\!\!$.3) is complimentary distributed with that of H\,{\sc i} , indicating possible evidence for the shock-cloud interaction, e.g. \citet{Sa16}. On the other hand, the 32$-$45 km s$^{-1}$ cloud appears to incomplete shell-like structure from the NW to the SW and shows an expanding structure in the velocity field. Figure \ref{gfigure_6} shows the position-velocity diagram of H\,{\sc i}. We can clearly see the expanding shell, the diameter of which is similar to that of the SNR. This trend indicates the physical association between the H\,{\sc i}  cloud and the SNR shell. We therefore conclude that both the 5$-$25 km s$^{-1}$ and 32$-$45 km s$^{-1}$ clouds are possibly associated with the SNR. Despite these results, the gamma-ray emission contours lie inside of the SNR, where there are no dense atomic and molecular clouds. At present, the 'illuminated cloud' scenario via the hadronic process is less favored in the SNR 3C 400.2. Further gamma-ray observation with finer angular resolution is needed to clarify the spatial correspondence between the interstellar gas and gamma rays. 

\section{Summary and Conclusions}
We have presented an investigation of the MM SNR 3C 400.2 using archival multi-wavelength data. The summary of the results and conclusions are listed below:
\begin{enumerate}

\item Using {\it Suzaku} data, we searched for RP from 3C 400.2 and determine X-ray properties of each region.  All regions require two thermal components, indicating that the plasma of 3C 400.2 contains a hard and soft temperature component. We find RP in the NE and SE regions. The selected NE and SE X-ray regions correspond to inside of the SNR radio shell, which is not much constrained by the dense atomic and molecular material. So, in these regions X-ray emission expanded into a void inside the SNR shell, thereby producing RP through the adiabatic cooling mechanism.

\item We report a detection of 3C 400.2 at the level of $\sim$5$\sigma$. Although the gamma-ray emission contours lie mostly inside of the SNR's shell, where there are no dense atomic and molecular clouds, a possible scenario for the gamma-ray emission from 3C 400.2 could be that dense atomic and molecular gas squeeze the SNR shell on the NE and NW sides, where the shell might get in contact with dense material and produce gamma rays through the hadronic scenario. Since the gamma-ray emission contours lie inside or on the radio shell of the SNR, the illuminating cloud scenario is less favored over the crushed cloud scenario. More data is needed to be able to show the hadronic emission model is the dominating model for 3C 400.2 over the leptonic models. 

\item We detected a new source, PS J1934.5$+$1845, at about 1$^{\circ}\!\!$.8 away from 3C 400.2 having a significance of $\sim$13$\sigma$. PS J1934.5$+$1845 was found to have a Radial Gaussian type extension with a radius of $\sim$0$^{\circ}\!\!$.61. We investigated this source as a part of our analysis due to the possibility that it might have been contributing to the gamma-ray emission of 3C 400.2. After fitting the extension and checking the TS value of 3C 400.2, we found out that the emission of PS J1934.5$+$1845 is not directly affecting the gamma-ray emission of 3C 400.2. The gamma-ray emission of PS J1934.5$+$1845 needs to be further investigated by modeling the SED in order to understand whether the dominating gamma-ray emission mechanism is leptonic or hadronic. 

\end{enumerate}

\section*{Acknowledgements}
We would like to thank Hideaki Matsumura for his useful comments on X-ray image analysis. T. Ergin thanks to the support of the Science Academy Young Scientists Program (BAGEP-2015). A. Sezer is supported by the Scientific and Technological Research Council of Turkey (T\"{U}B\.{I}TAK) through the B\.{I}DEB-2219 fellowship program. R.Yamazaki is supported in part by grant-in-aid from the Ministry of Education, Culture, Sports, Science and Technology (MEXT) of Japan, No. 15K05088. The Canadian Galactic Plane Survey (CGPS) is a Canadian project with international partners. The Dominion Radio Astrophysical Observatory is operated as a national facility by the National Research Council of Canada. The CGPS is supported by a grant from the Natural Sciences and Engineering Research Council of Canada. The NANTEN project is based on a reciprocal agreement between Nagoya University and the Carnegie Institution of Washington. We deeply acknowledge that the NANTEN project was realized by contributions from many Japanese public donors and companies.

$~$

$~$

{\it Facility}: {\it Suzaku}, {\it Fermi}, {\it NANTEN millimeter-wave telescope of Nagoya University}, {\it Dominion Radio Astrophysical Observatory}. 


\end{document}